# Broad-band dielectric response of $0.5Ba(Ti_{0.8}Zr_{0.2})O_3$–$0.5(Ba_{0.7}Ca_{0.3})TiO_3$ piezoceramics: soft and central mode behaviour

S. Kamba[1], E. Simon[1,2], V. Skoromets[1], V. Bovtun[1], M. Kempa[1], J. Pokorný[1],
M. Savinov[1], J. Koruza[3,4], B. Malič[3]

[1]*Institute of Physic, Czech Academy of Sciences, Prague, Czech Republic*
[2]*BCM College Kottayam, Kerala, India*
[3]*Institute Jozef Stefan, Ljubljana, Slovenia*
[4]*Technische Universität Darmstadt, Darmstadt, Germany*

kamba@fzu.cz

**Abstract**

Dielectric properties of $0.5Ba(Ti_{0.8}Zr_{0.2})O_3$–$0.5(Ba_{0.7}Ca_{0.3})TiO_3$ ceramics were probed in the frequency range from 10 Hz to 100 THz in a broad temperature range (10–900 K). Polar soft phonon observed in infrared spectra softens with cooling, however below 500 K its frequency becomes temperature independent. Simultaneously, a central mode activates in terahertz and microwave spectra; and it actually drives the ferroelectric phase transitions. Consequently, the phase transitions strongly resemble a crossover between the displacive and order–disorder type. The central mode vanishes below 200 K. The dielectric relaxation in the radiofrequency and microwave range anomalously broadens on cooling below $T_{C1}$ resulting in the nearly frequency independent dielectric loss below 200 K. This broadening comes from a broad frequency distribution of ferroelectric domain wall vibrations. Raman spectra reveal new phonons below 400 K, i.e. already 15 K above $T_{C1}$. Several weak modes are detected in the paraelectric phase up to 500 K in Raman spectra. Activation of these modes is ascribed to the presence of polar nanoclusters in the material.

1. **Introduction**

For years, there has been a stringent request of politicians and environmentalists for substitution of toxic lead-based piezoelectrics by some lead-free materials. Unfortunately, most of the lead-free piezoelectrics exhibit worse functional parameters (piezoelectric and electromechanical constants, thermal stability etc.) than those of lead-based materials.[1] In 2009, Liu and Ren [2] discovered high piezoelectric coefficients in a solid solution of $Ba(Ti_{0.8}Zr_{0.2})O_3$ and $(Ba_{0.7}Ca_{0.3})TiO_3$. They determined a phase diagram of $(1-x)Ba(Ti_{0.8}Zr_{0.2})O_3-x(Ba_{0.7}Ca_{0.3})TiO_3$ (subsequently abbreviated as BZT–$x$BCT) and reported a morphotropic phase boundary (MPB) between the tetragonal (T) and rhombohedral (R) phases near $x = 0.5$. Later studies questioned the term morphotropic phase boundary in this system because the boundary between T and R phase is not steep like in PZT or PMN-PT piezoelectrics. Regardless of existence or absence of MPB, BZT–$x$BCT ceramics with composition close to x = 0.5 exhibit the highest piezoelectric coefficient $d_{33}$~620 pC/N,[2] electric-field induced strain, and blocking force,[3] outperforming the most frequently used piezoelectric PZT. This discovery triggered intensive research on BZT–$x$BCT ceramics and thin films. Transmission electron microscopy [4] revealed coexistence of small T and R domains for compositions near x = 0.5; these domains allow easy polarization rotation between both polarization states in an external electric field, which explains the high $d_{33}$ coefficient. Curie–Weiss temperature $T_{C1}$ between the cubic (C) and T phase is around 370 K,[2] which is much lower than that in PZT. Just proximity of $T_{C1}$ to room (operation) temperature is probably responsible for the enhanced piezoelectric response of BZT–$x$BCT. The most studied composition BZT–0.5BCT exhibiting the highest $d_{33}$ shows two other structural and dielectric anomalies near $T_{C2} = 310$ and $T_{C3} = 280$ K. [5,6,7,8,9,10,11,12] There are still discussions in the literature whether these anomalies correspond to tetragonal-to-orthorhombic (T-to-O) and O-to-R phase transitions, respectively, or to the temperature range of the coexistence of T and R phases. Near 210 K Raman [6] and high-resolution X-ray [10] studies revealed another anomaly, which is probably correlated to the transition to a homogeneous R phase. Hao *et al*.[13] reported that the ceramics with the grain size below 1 μm exhibit relaxor ferroelectric behaviour with small piezo-coefficients, while the piezoresponse increases with the grain size and reaches its optimal values for the grains of 10 μm and larger.[13]

Up to now, BZT–$x$BCT lattice dynamics and its temperature dependence has been studied using only Raman spectroscopy.[6,14] The ferroelectric soft mode can be Raman active only in the ferroelectric phase, while it should be infrared (IR) active in both paraelectric and ferroelectric phases. For this reason we have performed IR and terahertz (THz) studies in the broad temperature region from 10 to 900 K as well as new

Raman measurements. Since pure BaTiO$_3$, BZT and BCT exhibit a broad dielectric relaxation (called central mode) below the soft mode frequency,[15,16,17] we have also performed microwave and radio-frequency dielectric measurements. Temperature behaviour of the soft and central mode is reported in detail.

2. **Experimental**

The BZT–0.5BCT ceramics were prepared from BaCO$_3$ and CaCO$_3$ (both 99.95 %, Alfa, Germany), TiO$_2$ (99.8 %, Alfa, Germany) and ZrO$_2$ (99.8 %, Tosoh, Japan). The reagents in the stoichiometric ratio were homogenized, milled in a planetary mill (Y$_2$O$_3$–ZrO$_2$ milling bodies, acetone, 4 h, 200 rpm) and calcined at 1350 °C for 4 h. The as-calcined powder was again milled for 4 h under identical conditions and dried afterwards. The powder was compacted uniaxially at 100 MPa and then isostatically at 200 MPa. The powder compacts with the diameter of 8 mm were sintered at 1450 °C for 4 h with the heating and cooling rates of 5 °C/min in the air atmosphere.

The density of the ceramics determined by the Archimedes method was of 5.57 g/cm$^3$, which constitutes 93.6 % of the theoretical density. According to XRD data, the material was a single phase perovskite. The average grain size assessed by a microstructural analysis of the samples was of 12.5 µm.

Low-frequency (10 Hz–1 MHz) complex dielectric permittivity measurements were performed between 10 and 480 K using NOVOCONTROL Alpha-A High Performance Frequency Analyzer (a Leybold cryostat was used below 300 K and a home-made furnace above room temperature). A gold electrode structure was sputtered on the polished ceramic plate with the thickness of 350 µm and 5 mm in diameter.

High-frequency (1 MHz–1.8 GHz) dielectric measurements were performed using a computer controlled dielectric spectrometer equipped with an Agilent 4291B impedance analyzer, Novocontrol BDS 2100 coaxial sample cell and Sigma System M18 temperature chamber (operating range of 100–500 K). Dielectric parameters were calculated taking into account the electromagnetic field distribution in the rod-shaped samples (diameter of 1 mm and length of 7 mm) with Au sputtered electrodes.

Time-domain THz spectroscopic experiments were carried out using a custom-made spectrometer powered by a Ti:sapphire femtosecond laser. The system is based on coherent generation and subsequent coherent detection of ultra-short THz transients.[18] The detection scheme is realized on an electro-optic sampling of the electric field of the transients within a 1 mm-thick (110)-oriented ZnTe crystal as a sensor. This allows us to measure time profile of the THz transients transmitted through a studied sample. For these measurements, a plane-parallel sample plate with a diameter of 6 mm and thickness of 77 μm was prepared. For the THz and IR measurements at temperatures below 300 K we put the samples into an optical cryostat (Optistat, Oxford Instruments). For higher temperatures we used a commercial high-temperature cell (SPECAC P/N 5850).

Near-normal-incidence IR reflectivity spectra were measured using a Fourier-transform IR spectrometer Bruker IFS 113v. The spectral range of the low-temperature IR measurements was limited by the transparency of polyethylene windows (up to 650 cm$^{-1}$) in cryostat, whereas the spectral range above room temperature was up to 3000 cm$^{-1}$. At high temperatures, two deuterated triglycine sulfate detectors were used for far and middle IR region. At low temperatures, a liquid-He-cooled Si bolometer operating at 1.6 K served as a far-IR detector. The same cryostat and high-temperature cell as in the THz experiments, were used for the low- and high-temperature IR measurements, respectively.

Raman scattering experiments were performed in the frequency range of 100–1000 cm$^{-1}$. Unpolarized spectra excited with a 514.5 nm line of an Ar laser were recorded in a back-scattering geometry with a Renishaw RM 1000 micro-Raman spectrometer equipped with edge filters. The sample was put in a Linkam THMS600 cell allowing spectra acquisition in a broad temperature interval from 80 to 600 K. The laser beam with a power of ~10 mW was focused on a ~5 μm spot. The spectra were corrected to the Bose–Einstein temperature factor and fitted to a sum of damped harmonic oscillators using an in-house software.

3. **Results and discussion**

In Figure 1 we show temperature dependences of the relative complex dielectric permittivity $\varepsilon^*(T)=\varepsilon'(T) - i\varepsilon''(T)$ probed at frequencies between 10 Hz and 200 GHz. In the literature, $\varepsilon^*(T)$ has been reported mostly at a single frequency in the kHz region,[2,13] or in a narrow radio-frequency region.[9,19] In this work we first show $\varepsilon^*(T)$ obtained at 10 Hz and at frequencies above 1 MHz. Besides a huge anomaly at $T_{C1} = 375$ K corresponding to the ferroelectric phase transition, other two anomalies can be clearly seen near $T_{C2} = 285$ K and $T_{C3} = 236$ K at frequencies below 300 MHz. They most probably correspond to the structural phase transitions to the R and O phases. Our $T_{C2}$ and $T_{C3}$ are slightly lower than those reported in the literature. This might be caused by slight deviationsin the composition or microstructure of BZT–0.5BCT ceramics. Note that the anomalies are the most pronounced at 10 Hz, where two sharp peaks are clearly seen in temperature dependence of dielectric loss. Other authors reported dielectric data mainly in kHz region, where the sensitivity is lower. $\varepsilon'(T)$ probed at 200 GHz exhibits a maximum shifted ~100 K above $T_{C1}$ (this will be explained below). Dielectric properties were not measurable at 200 GHz between 380 and 475 K because the sample was completely opaque in the THz region at these temperatures.

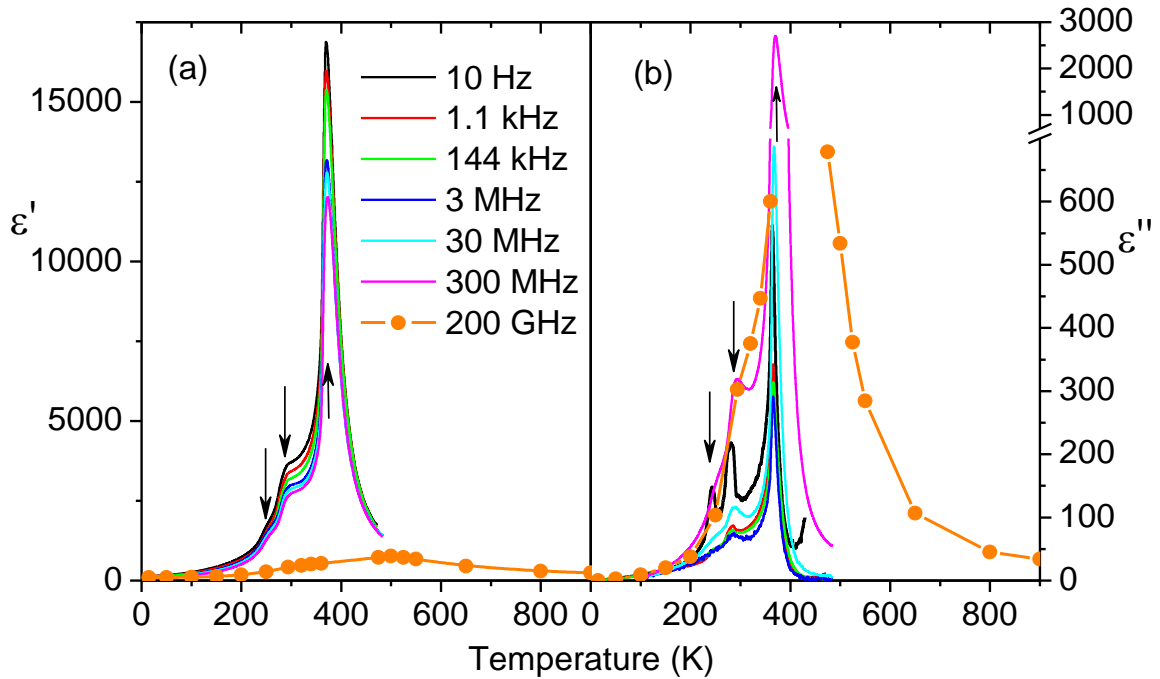

Figure 1.Temperature dependence of (a) the dielectric permittivity and (b) loss in BZT–0.5BCT ceramics. Temperatures of dielectric anomalies are marked by arrows. Note the change of scale in dielectric loss $\varepsilon''$.

Frequency spectra $\varepsilon^*(\omega)$ are shown in Figure 2. There is a gap in the data between 1.8 GHz and 100 GHz, nevertheless one can expect that there is a relaxation (central mode) in this frequency range because dielectric loss $\varepsilon''(\omega)$ spectra tend to have a peak here and THz permittivity $\varepsilon'(\omega)$ spectra have lower values than microwave $\varepsilon'(\omega)$. The microwave relaxation weakens towards low temperatures, but does not disappear down to 100 K. Its mean frequency does not change essentially with temperature and stay in GHz range in the whole studied temperature interval. Another relaxation is observed at low frequencies (below 1 MHz); note the increase in $\varepsilon''(\omega)$ spectra on frequency decrease. This relaxation could be related to the defects and other inhomogeneities in the ceramic, which contribute to a conductivity.

At low temperatures, we observe the linear dependence of $\varepsilon'(\omega)$ and nearly frequency independent dielectric loss $\varepsilon''(\omega)$ in a relatively broad frequency range below 1 GHz. A similar behaviour is known in relaxor ferroelectrics, where the dynamics of polar nonoclusters with a very broad distribution of relaxation times is responsible for frequency independent dielectric loss below freezing temperature.[20] In our case, the frequency range of constant loss is narrower than in relaxor ferroelectric PMN [21] and moreover, it is observed in ferroelectric phases without any polar nanoclusters. We could suppose that this dielectric relaxation is related to the complicated domain structure in BZT-0.5BCT ceramics near polymorphic phase transitions (PPT) and comes, for instance, from vibrations of ferroelectric domain walls. Since the domains are very small near PPT [4] and there is a large distribution of their sizes, there is also a broad distribution of the domain wall vibration frequencies which will even broaden on cooling. Another origin of the constant loss might be attributed to some peculiarities of the mesoscopic structure of the grain boundaries in BZT-BCT ceramics. Let us note that similar behaviour was observed in KNN ceramics [22], more pronouncing in the small-grain ones, and in fine-grain PMN-PT ceramics with a fine domains in the ferroelectric phase. [23]

Above 100 GHz and at low temperatures (below 250 K), $\varepsilon''(\omega)$ exhibits increase with frequency. This feature corresponds to a low-frequency wing of the soft phonon.

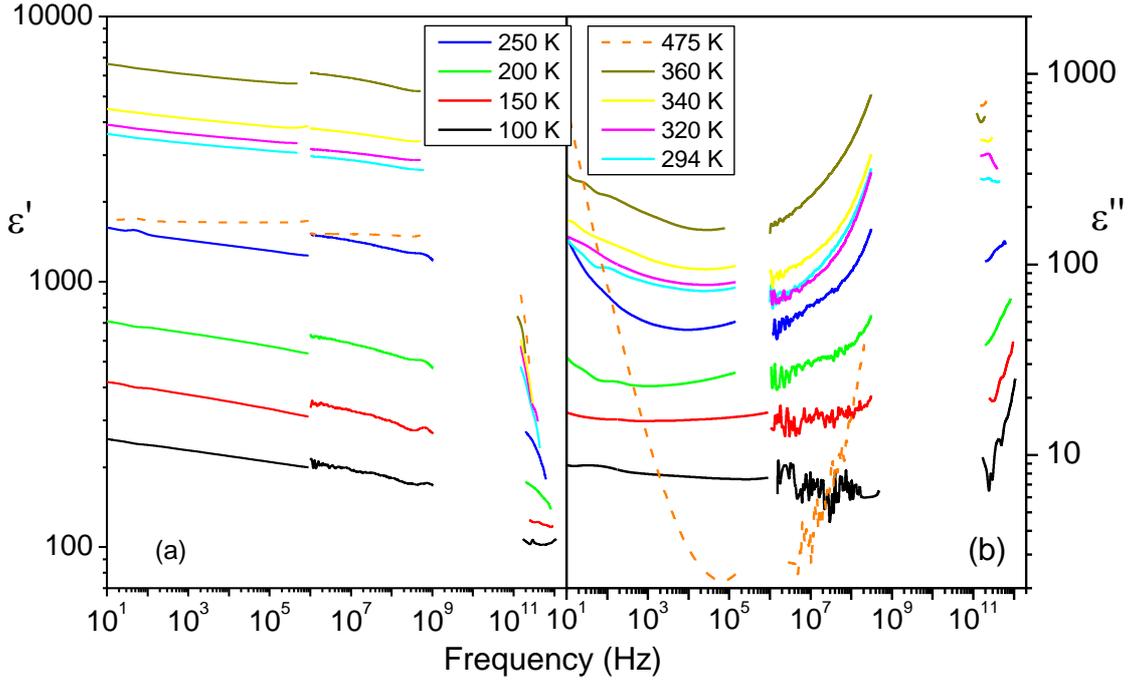

Figure 2. Frequency dependence of (a) the dielectric permittivity and (b) loss in BZT–0.5BCT ceramics plotted at various temperatures. Discontinuities in the spectra come from using three different experimental techniques.

Temperature behaviour of IR reflectivity spectra together with recalculated THz data are shown in Figure 3. One can see sharpening of all phonons on cooling due to lowering of the phonon damping. Intensity of the lowest frequency band decreases on cooling due to a weaker contribution of the central mode to the THz and microwave permittivity. The IR reflectivity spectra with a complex dielectric response obtained from the time-domain THz measurements were fitted simultaneously using a standard factorized form of the complex permittivity.[17] Results of the fits are presented in Figure 4. Frequencies of the peaks in the $\varepsilon''(\omega)$ spectra correspond to transversal phonon frequencies $\omega_{TO}$. Below 10 cm$^{-1}$ one can see an additional broad peak which was fitted by an overdamped oscillator. This feature is actually the high-frequency wing of the central mode observed in the microwave dielectric spectra (Figure 2). Figure 5 shows temperature dependence of all detected phonon frequencies. The mode near 300 cm$^{-1}$ exhibits an unusual behaviour: it disappears below 600 K. This weak mode probably comes from a multiphonon absorption or from an imperfect correction of the emission spectra. Besides this mode we detect six phonons in the cubic paraelectric phase, although only three polar phonons are allowed in the ideal cubic perovskite $ABO_3$.[24] Twice as higher number of the observed phonons is caused by two kinds of cations sitting in the A (Ba,Ca) and B (Ti,Zr) perovskite sites. The same number of phonons was observed in cubic phase of BZT ceramics.[16] Since most of the phonons have rather large damping above room temperature, one can detect newly activated phonons only below 175 K, i.e. far below the ferroelectric phase transition at 385 K. However, the new phonons should be activated already at higher temperatures due to symmetry lowering below $T_{C1}$. These phonons are rather weak and they are probably screened by the other strong and broad phonons between 200 and 385 K. Seven and twelve IR active modes are allowed in the R and O phase of $BaTiO_3$.[24] In Figure 5 we can detect ten modes in the R phase below 175 K. A higher number of the modes than that allowed from the factor group analysis is again, as in the paraelectric phase, caused by the double occupancy of the A and B perovskite sites.

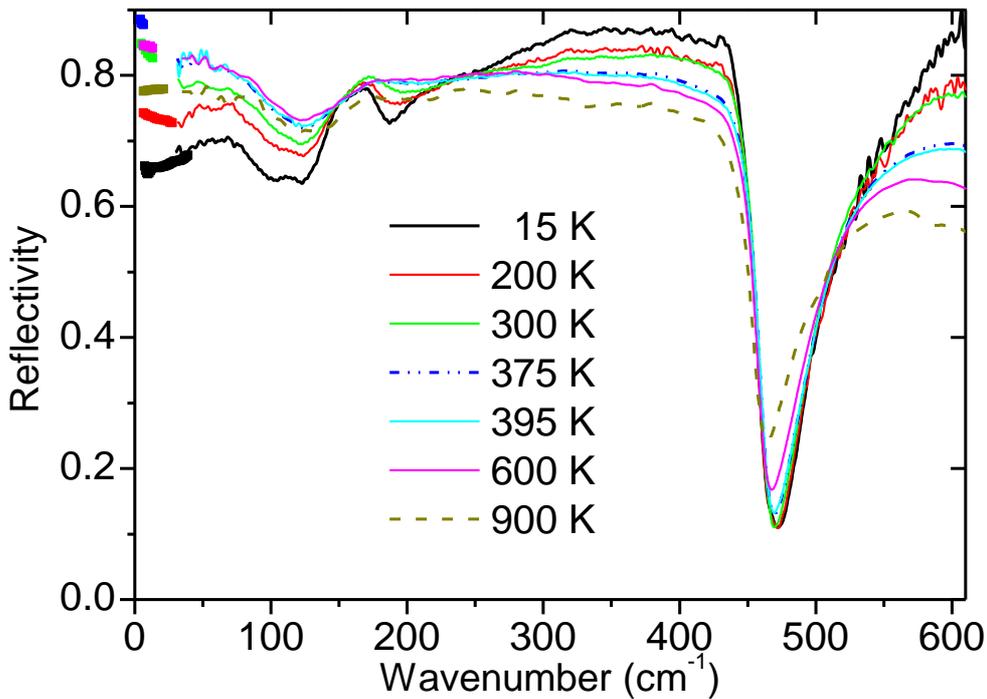

Figure 3. Far-IR reflectivity spectra of BZT–0.5BCT ceramics plotted at various temperatures. Data below 30 cm$^{-1}$ were calculated from the complex THz dielectric spectra.

The central mode is activated in the spectra not only near the critical temperatures, but it is detectable already from 200 to 500 K. This gives evidence about a crossover (occurring between 500 and 200 K) from the displacive to the order–disorder mechanism of the phase transition. Note that, on the one hand, the central mode was also detected in the spectra of pure $BaTiO_3$ ceramics in the whole investigated temperature range from 10 to 830 K,[16,17] and BZT exhibits relaxor ferroelectric-like behaviour for more than 30 % of Zr content, but its relaxation frequency slows down on cooling according to the Arrhenius behaviour. [16,17] On the other hand, BCT undergoes the first-order ferroelectric phase transition to the tetragonal phase, although its behaviour is very similar to pure $BaTiO_3$: the soft mode is overdamped at all temperatures [15] and its behaviour resembles that of $BaTiO_3$. In BCT the phase transitions to the O and R phases are strongly supressed; and if they occur then only at temperatures below 40 K. [15]

As was mentioned above, $\varepsilon'(T)$ measured at 200 GHz exhibits maximum at ~500 K, shifted ~100 K above $T_{C1}$ (Figure 1). It reminds the shift of $\varepsilon'(T)$ maximum towards Burns temperature in classical relaxor ferroelectrics [21,25] and BZT [16]. Note that the central mode is activated in the spectra at the same temperature and we will show below that $TO_2$ polar phonon is also activated in Raman spectra at 500 K. This constitutes an evidence of appearance of some polar clusters near this temperature. Therefore, $T_B$ = ~500 K can be considered as the Burns temperature in BZT–BCT. This is in a good agreement with the value of $T_B$ = ~450 K in BZT [17, 26].

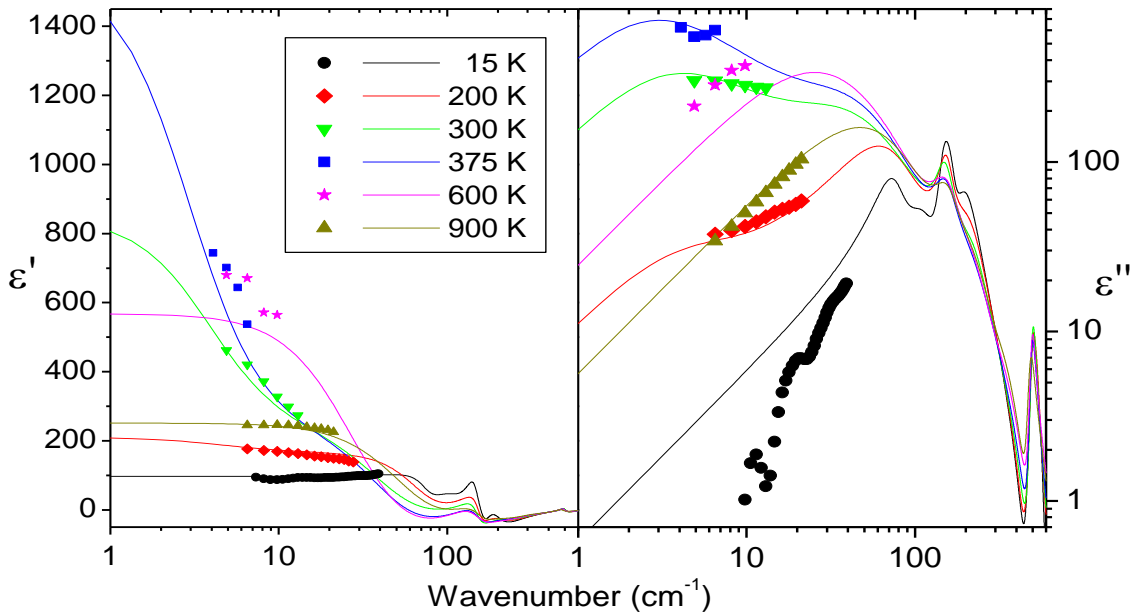

Figure 4. Complex dielectric spectra of BZT–0.5BCT obtained from the fits of the THz and IR spectra. Symbols at low frequencies are experimental THz spectra.

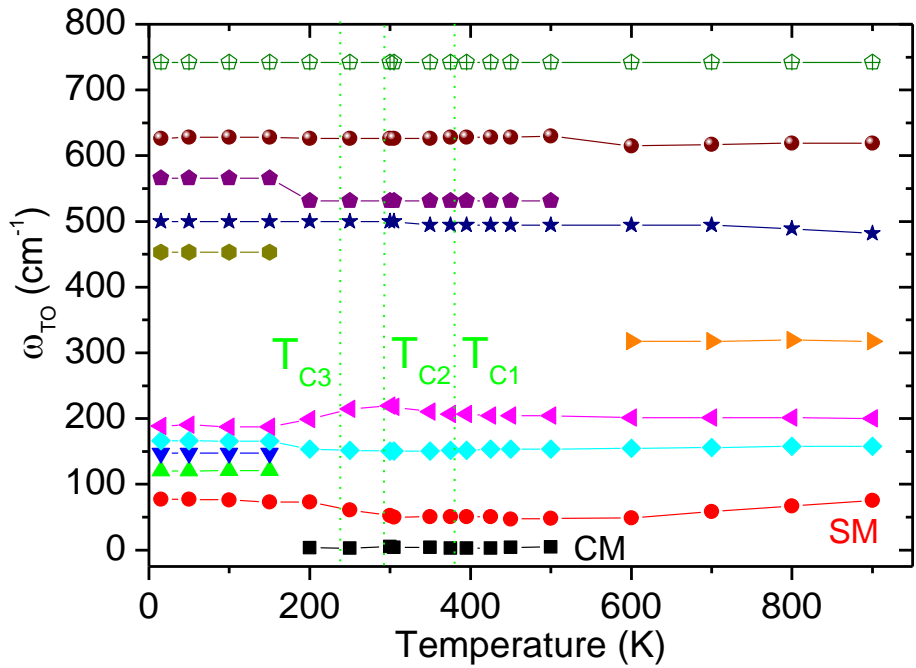

Figure 5. Temperature dependence of transverse optical phonon frequencies obtained from the fits. Soft and central modes are marked as SM and CM, respectively.

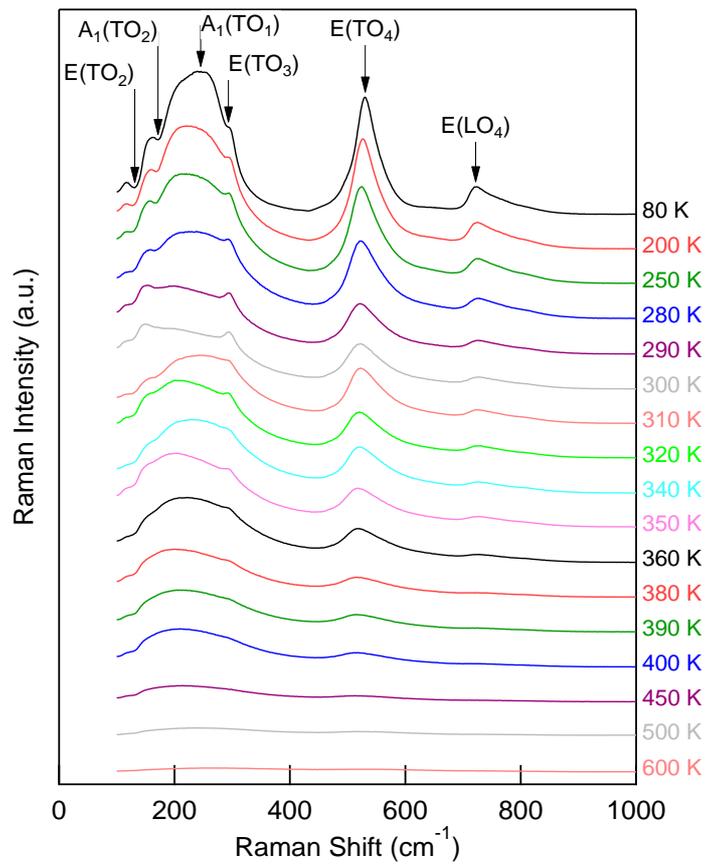

Figure 6. Unpolarized Raman scattering spectra of the BZT–0.5BCT ceramics recorded at various temperatures. Symmetries of the modes are assigned.

In the cubic phase, all phonons should be Raman inactive. Far above $T_{C1}$ we observed two weak and broad asymmetric bands near 200 and 550 cm$^{-1}$ (Figure 6), which actually consist of four modes (Figure 7). Similar Raman peaks are present practically in all cubic perovskites and usually have their origin in Raman scattering of the second order. Below 500 K, a new mode corresponding to $E(TO_2)$ appears near 160 cm$^{-1}$. It confirms local breaks of the cubic symmetry already above $T_{C1}$. The same mode was observed previously in published Raman spectra of BZT–0.5BCT ceramics even up to 546 K.[14] Signature of the local regions with a broken symmetry far above $T_C$ was discovered also in BaTiO$_3$ by measurements of the second harmonic signal. [27] Two new Raman peaks appear near 300 cm$^{-1}$ and 720 cm$^{-1}$ at 380 K, i.e. just below $T_{C1}$. These are $E(TO_3)$ and $E(LO_4)$ modes, respectively. A sharp mode near 530 cm$^{-1}$ corresponds to $E(TO_4)$ mode. A minimum observed near 190 cm$^{-1}$ is ascribed to $A_1(TO_2)$ mode. The soft $E(TO_1)$ mode is not detectable because it lies below our frequency range, i.e. below 100 cm$^{-1}$. Only $A_1$ component of $TO_1$ mode can be observed near 250 cm$^{-1}$. Most of the Raman modes do not exhibit any dramatic frequency changes near $T_{C2}$ and $T_{C3}$. Only the mode observed near 290 cm$^{-1}$ exhibits some weak softening near $T_{C3}$. Stability of the remaining modes gives evidence about small differences between tetragonal, orthorhombic and rhombohedral phases.

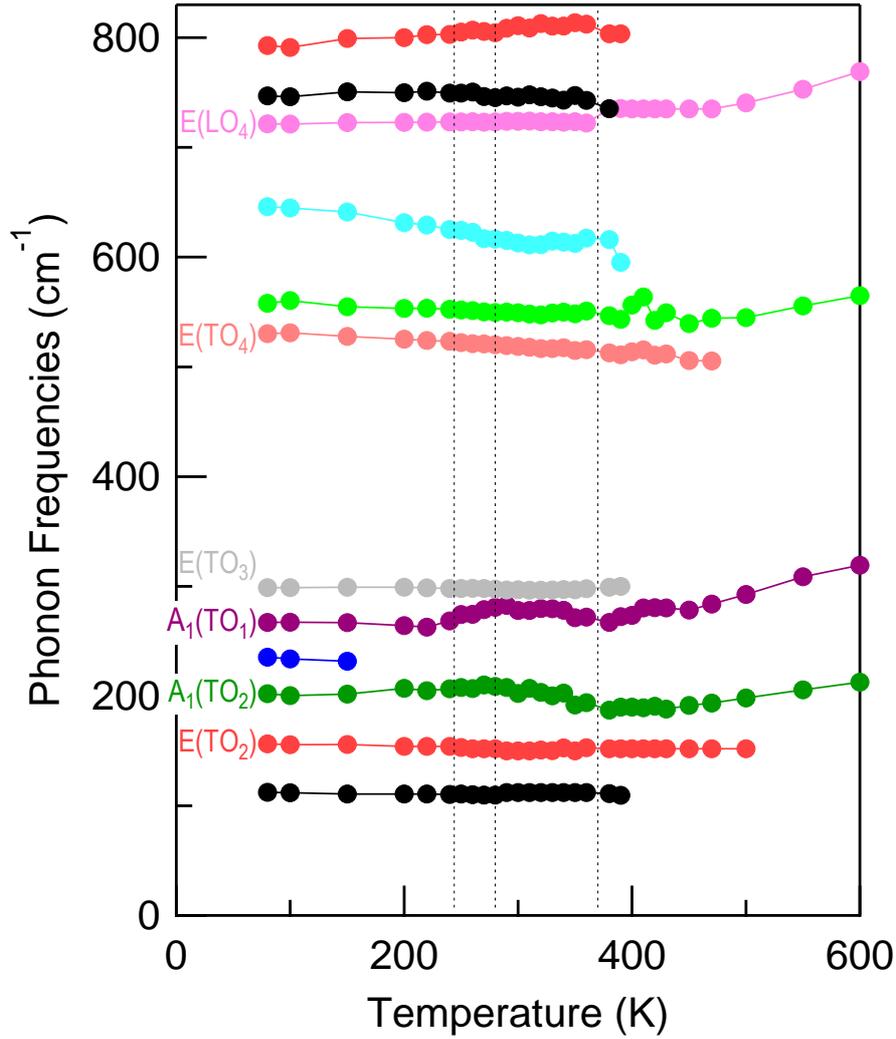

Figure 7. Temperature dependence of the Raman active phonon frequencies in the BZT–0.5BCT ceramics.

## 4. Conclusion

We studied vibrations of the crystal lattice and the domain walls in BZT–0.5BCT ceramics in a broad temperature and frequency region. In the paraelectric phase, IR spectra revealed a number of phonons twice higher than allowed from the factor group analysis. This is caused by splitting of the allowed phonons because each A and B perovskite sites contain two kinds of the cations with different masses. The phonons split in the ferroelectric phase due to symmetry lowering. The lowest frequency phonon softens on cooling, however below 500 K it becomes temperature independent. Simultaneously, in the THz and microwave spectra we observe the activation of the central mode, which actually drives the phase transitions. The central mode has a frequency in the range of 10–100 GHz and it markedly weakens below 200 K, i.e. deeply below the phase transition to the rhombohedral phase. Below 200 K, a dielectric relaxation develops in the radiofrequency and microwave spectra. It has a broad distribution of relaxation frequencies and, therefore, it leads to a nearly frequency independent dielectric loss between 10 Hz and 1 GHz. Domain wall vibrations could be at the origin of this behaviour, which resembles dynamics of polar nanoclusters in relaxor ferroelectrics. The temperature dependence of the dielectric permittivity and loss at low frequencies clearly revealed three anomalies giving an evidence of successive phase transitions from the cubic to the tetragonal, orthorhombic and rhombohedral phases. Raman scattering spectra confirmed a presence of polar nanoclusters in the paraelectric phase at least up to 500 K. The temperature dependence of the permittivity measured at 200 GHz exhibits a peak near 500 K, giving additional evidence about the presence of the polar nanoregions in the paraelectric phase. The Burns temperature in BZT-BCT ceramics is estimated to be slightly above 500 K.

**Acknowledgement**

This work was supported by the Czech Science Foundation, Project 15–08389S and by MŠMT project LD15014. BM and JK acknowledge the support of the Slovenian Research Agency (programme P2-0105).

---